\documentclass{entcs}
\usepackage{prentcsmacro}

\usepackage{amssymb}
\usepackage{amsfonts}
\usepackage{latexsym}
\usepackage{epsfig}
\usepackage{diagrams}
\usepackage{verbatim}
\usepackage{color}

\usepackage{pstricks}

 \newarrow{Inj}>--->

\usepackage{xy}
\xyoption{all}

\newcommand{\four}{{\rm I}\hspace{-0.4mm}{\rm V}}

\newcommand{\bit}{\begin{itemize}}
\newcommand{\eit}{\end{itemize}\par\noindent}
\newcommand{\ben}{\begin{enumerate}}
\newcommand{\een}{\end{enumerate}\par\noindent}
\newcommand{\beq}{\begin{equation}}
\newcommand{\eeq}{\end{equation}\par\noindent}
\newcommand{\beqa}{\begin{eqnarray*}}
\newcommand{\eeqa}{\end{eqnarray*}\par\noindent}
\newcommand{\beqn}{\begin{eqnarray}}
\newcommand{\eeqn}{\end{eqnarray}\par\noindent}

\newcommand{\bpf}{\noindent {\bf Proof:} }
\def\endproof{\hfill$\Box$}

\def\II{{\rm I}}

\newcommand{\ket}[1]{|#1\rangle}

\newcommand{\blk}{\color{black}}

\newarrow{Mo}|--->

\def\lastname{Coecke, Edwards and Spekkens}

\begin{document}
\begin{frontmatter}

\title{Phase groups and the origin\\ of non-locality for qubits}
\author{Bob Coecke, Bill Edwards and Robert W. Spekkens}
\address{Oxford University Computing Laboratory\\ \ \\ and\\ \ \\ Perimeter Institute for Theoretical Physics}
\thanks{We thank Joachim Kock for carefully proofreading this manuscript and providing very useful feedback. This work is supported by EPSRC Advanced Research Fellowship EP/D072786/1, by US Office of
Naval Research (ONR) Grant N00014-09-1-0248, by EU-FP6-FET STREP QICS and by a Large Grant of the Foundational Questions Institute.}
\maketitle

\begin{abstract}
We describe a general framework in which we can precisely compare the structures of quantum-like theories which may initially be formulated in quite different mathematical terms. We then use this framework to compare two theories: quantum mechanics restricted to qubit stabiliser states and operations, and Spekkens's toy theory. We discover that viewed within our framework these theories are very similar, but differ in one key aspect - a four element group we term the \emph{phase group} which emerges naturally within our framework. In the case of the stabiliser theory this group is $Z_4$ while for Spekkens's toy theory the group is $Z_2 \times Z_2$. We further show that the structure of this group is intimately involved in a key physical difference between the theories: whether or not they can be modelled by a local hidden variable theory. This is done by establishing a connection between the phase group, and an abstract notion of GHZ state correlations. We go on to formulate precisely how the stabiliser theory and toy theory are `similar' by defining a notion of `mutually unbiased qubit theory', noting that all such theories have four element phase groups. Since $Z_4$ and $Z_2\times Z_2$ are the only such groups we conclude that the GHZ correlations in this type of theory can only take two forms, exactly those appearing in the stabiliser theory and in Spekkens's toy theory. The results point at a classification of local/non-local behaviours by finite Abelian groups, extending beyond qubits to finitary theories  whose observables are all mutually unbiased.
\end{abstract}
\end{frontmatter}

\section{Introduction}
Much interest recently has focused on picking out the key features of quantum mechanics which make it special (for example, incompatible observables, non-locality, computational speed-up, no-cloning), investigating the relationships between these features, and identifying the mathematical aspects of the theory which embody these physical features. Quantum-like theories have been constructed, which display certain features of quantum mechanics but not others, allowing us to see which of these features are interlinked, and which are essentially independent.

These theories had diverse motivations and are expressed in a range of mathematical forms. Quantum mechanics uses Hilbert space. Another theory which has recently attracted much interest \cite{Spekkens} employs subsets of certain sets to represent states, and relations between these sets to represent the operations of the theory. Other quantum-like theories use quite different mathematical formalisms again \cite{NLB,BBLW08}. The task of comparing these theories would be simplified if we had a single mathematical framework in which they could all be expressed. We could then pinpoint aspects of the framework where theories differed, and identify these aspects with differing physical features of the theories.

This paper will outline such a framework, developed in \cite{AC,Selinger,CP,CPV,CD}, and then use it to analyse some key examples. In this case the physical property we will be interested in is non-locality. To this end we extend the existing framework to encompass an abstract definition of GHZ state, and a corresponding notion of correlations.

What is nonlocality? The name tells us that ``it's \underline{not} locality.''  The technical definition tells us that ``there is \underline{no} local hidden variable theory.''  By Bell's Theorem this means that ``some inequality is \underline{not} satisfied.''  All this tells us what nonlocality is not, but what actually ``is'' nonlocality?  It is our goal in this paper to identify the \em piece of structure \em of Hilbert space quantum mechanics that \em generates non-locality\em.

To this end we will use our framework to analyse two theories which make very similar predictions, but differ principally in that one is local and the other is non-local. We will express both standard quantum mechanics, and a quantum-like toy theory proposed by one of the authors \cite{Spekkens}, called Spekkens's toy theory, in the unifying framework. The toy theory replicates many features of QM (e.g. incompatible observables, teleportation, no-cloning), but it is essentially a local hidden variable theory, and so it lacks other typically quantum features, specifically violation of Bell inequalities, and other `non-local' behaviour. We will identify a key piece of the structure of the unifying framework where QM and the Spekkens's toy theory differ (an Abelian group we term the \emph{phase group}). Furthermore we will show explicitly that it is this piece of structure which in the QM case facilitates a `no-go theorem' which rules out a hidden variable interpretation. Conversely, in the toy theory case, the phase group does not allow construction of such a no-go theorem. We speculate that this key piece of structure is responsible for the locality/non-locality of any quantum-like theory that our framework is capable of encompasing.

\section{The framework: Dagger compositional theories}

To make a comparison between qubit stabiliser formalism and Spekkens's toy theory we need a framework with concepts that are sufficiently general to accommodate both of them.  In particular, we need to be able to speak about GHZ states and observables for theories other than Hilbert space quantum mechanics.  Such a general account of physical theories was initiated by Abramsky  and one of the authors in \cite{AC}, and further developed by several others \cite{Selinger,CP,CPV,CD}. We refer the reader for physicist friendly introductions and tutorials on symmetric monoidal categories to \cite{Cats,LNPBS} and \cite{LNPCP,LNPAT} respectively.

The operational foundation for these structures is as follows -- detailed discussions are in \cite{Cats,LNPCP}.  \em Systems \em are represented by their names $A, B, C, \dots$ \em Processes \em (or \em operations\em) are represented by arrows $A\rTo^{f}B$ or $f:A\to B$ indicating the initial system $A$ and the resulting system $B$.  \em States \em are special arrows $\psi:\II\to A$ where $\II$ is the `unspecified' system.  We can sequentially compose processes if the intermediate systems match i.e.~$A\rTo^{h\circ g}C \ = \ A\rTo^{g}B\rTo^{h}C$.  There also are processes which leave the system invariant: $A\rTo^{1_A}A$.  Compound systems are denoted $A\otimes B$ and separate processes thereon $A\otimes C\rTo^{f\otimes g} B\otimes D$.  We refer to the arrows $\II\rTo^{s}\II$ as \em numbers\em. In addition we assume that each process $A\rTo^{f}B$ comes equipped with an adjoint process  $B\rTo^{f^\dagger}A$.   The precise mathematical notion which accounts for how \emph{sequential composition}, denoted $\circ$, and the \em tensor\em, denoted $\otimes$, interact is that of a dagger symmetric monoidal category. 

\begin{definition}
A \em dagger symmetric monoidal category \em ${\bf C}$ is a category with a bifunctor $-\otimes-:{\bf C}\times{\bf C}\to{\bf C}$, associativity, unit and symmetry natural isomorphisms, subject to the usual coherence conditions, and a contravariant involutive functor ${-}^\dagger:{\bf C}\to{\bf C}$ which coherently preserves the monoidal structure.
An arrow $f:A\to B$ is \em unitary \em if we have $f^\dagger\circ f=1_A$ and $f\circ f^\dagger=1_B$.
We assume associativity, unit and symmetry natural isomorphisms to be unitary.
\end{definition}  

Each such dagger symmetric monoidal category moreover admits a purely diagrammatic calculus \cite{Cats,LNPCP,Selinger,Selinger2}, for example:
\begin{center}
\qquad$f$\ $\equiv$\ \raisebox{-0.85cm}{\epsfig{figure=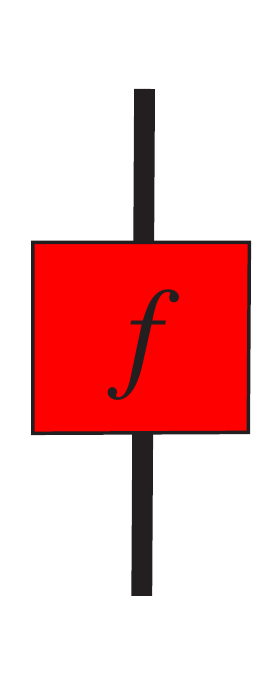,width=20pt}}\qquad\quad $   1_A$\ $\equiv$\ \raisebox{-0.85cm}{\epsfig{figure=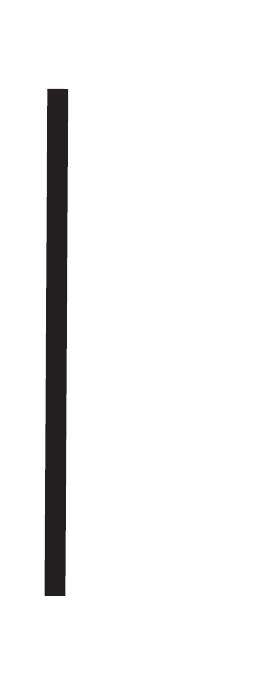,width=20pt}}
\quad\qquad
$g\circ f$\ $\equiv$\ \raisebox{-0.85cm}{\epsfig{figure=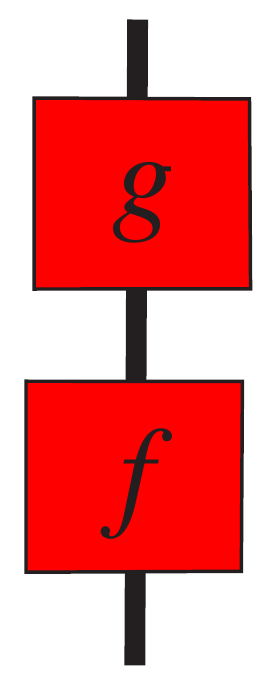,width=20pt}}\quad\qquad $f\otimes g$\ $\equiv$\    \raisebox{-0.55cm}{\epsfig{figure=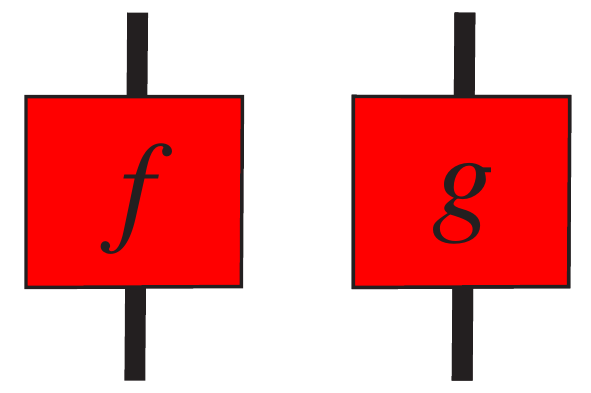,width=50pt}}
\end{center}
These diagrams are not merely denotation but are truly equivalent to the algebraic symbolic presentation in the  following sense.

\begin{theorem}{\rm\cite{Selinger}}
An equation expressible in the language of dagger symmetric monoidal categories is provable from the axioms of a dagger symmetric monoidal category if and only if it is derivable in the corresponding diagrammatic calculus.
\end{theorem}

\begin{definition}
A \em dagger compositional theory\em, or in short, \em $\dagger$C-theory\em, is a dagger symmetric monoidal category in which we interpret objects as systems, morphisms as processes, with states and effects as the particular cases arising from the unit object, composition as performing one process after the other, and the tensor as composing systems and parallelling processes.
\end{definition}

\begin{example}\label{FHilbexample}
In the $\dagger$C-theory ${\bf FHilb}$ the objects are finite dimensional Hilbert spaces, the arrows are linear maps, the tensor is the tensor product, and the dagger is the linear algebraic adjoint. 
States are of the form
\[
|\psi\rangle:\mathbb{C}\to{\cal H}::1\mapsto\psi\,,
\]
and hence  correspond to vectors, and the numbers are of the form
\[
\langle c\rangle:\mathbb{C}\to\mathbb{C}::1\mapsto c\,,
\]
and hence  correspond to complex numbers.  One can interpret the morphisms in \textbf{FHilb} as pure quantum processes with
postselection, that is, all the vectors, dual vectors and linear maps (because
postselected logic gate teleportation allows us to produce any linear map up to a
probabilistic weight).\footnote{In the language of quantum information theory,
these include the pure density operators, the rank-1 effects, and the completely
positive maps with a single Kraus operator, as well as unnormalized versions
thereof.}
\end{example}

\begin{example}\label{FRelexample}
In the $\dagger$C-theory  ${\bf FRel}$, the objects are finite sets, the arrows are relations, the tensor is the cartesian product, the dagger is the relational converse.  The identity object is the single element set $\{*\}$.  Now states are of the form
\[
|r\rangle:\{*\}\!\to X::*\mapsto Y\!\subseteq\! X\,,
\]
and hence correspond to subsets,  and the numbers are of the form
\[
\langle b\rangle:\{*\}\!\to\!\{*\}\!::*\mapsto \emptyset\mbox{ or }*\,,
\]
and hence correspond to the booleans.  One can interpret these relational operations as  `possibilistic' (classical) processes.
\end{example}

\begin{example}[From vectors to rays] \label{ex:wp}
The states of the $\dagger$C-theory ${\bf FHilb}$ as defined above are vectors in a Hilbert space, not one-dimensional subspaces.  In other words, it contains physically redundant `global phases'. One way to eliminate these global phases is by considering equivalence classes of linear maps that are equal up to a global phase. Another way applies to arbitrary $\dagger$C-theories:
\end{example}

\begin{definition}\cite{deLL} [\em${\cal W}$-construction\em]
Given a $\dagger$C-theory ${\bf C}$ we define a $\dagger$C-theory ${\cal W}{\bf C}$ to have the same objects as ${\bf C}$, with ${\cal W}{\bf C}(A,B):=\{f\otimes f^\dagger\mid f\in {\bf C}(A,B)\}$, and with $(f\circ g)\otimes (f\circ g)^\dagger$ as the composite of $f\otimes f^\dagger$ and $g\otimes g^\dagger$.  For $f$ a morphism in ${\bf C}$ we set  ${\cal W}f:=f\otimes f^\dagger$ for the corresponding morphism in ${\cal W}{\bf C}$.
\end{definition}

\begin{example}[From vectors to rays continued] This ${\cal W}$-construction has the added advantage that expressions of the form $\langle\psi|\phi\rangle:=\psi^\dagger\circ\phi$, after application of the ${\cal W}$-construction, become $|\langle\psi|\phi\rangle|^2=(\psi^\dagger\circ\phi)^\dagger\circ(\psi^\dagger\circ\phi)$, that is, transition probabilities according to the Born rule.  For states in ${\bf FHilb}$, applying the ${\cal W}$-construction essentially boils down to the same thing as passing from kets $|\psi\rangle$ to projectors $|\psi\rangle\langle\psi|$ in the density matrix formulation.  The numbers in ${\cal W}{\bf FHilb}$ are positive real numbers, which we interpret as probabilistic weights.  We have ${\cal W}({\cal W}{\bf FHilb})\simeq{\cal W}{\bf FHilb}$, and ${\cal W}{\bf FRel}\simeq{\bf FRel}$.
\end{example}

\begin{definition}\cite{deLL}
A $\dagger$C-theory ${\bf C}$ is  \em without global phases \em iff ${\cal W}{\bf C}\simeq{\bf C}$
\end{definition}

Below, we will always assume that we have eliminated the global phases from ${\bf FHilb}$, even when we write ${\bf FHilb}$.  Hence the numbers in this category are the positive reals, which we interpret as probabilistic weights. Inner-products then provide the correct transition probabilities according to the Born rule.  More generally, we will interpret the numbers in $\dagger$C-theories as probabilistic weights and inner-products $\langle -|-\rangle:=(-)^\dagger\circ(-)$ as transition probabilities.

\section{Key features of the $\dagger$C-theory framework}\label{keyfeaturessection}

\subsection{Observables in $\dagger$C-theories}\label{observablesection}

In this section we explain how the usual notion of a non-degenerate observable can be generalised from Hilbert spaces to other $\dagger$C-theories.

\begin{definition}\label{def:observable}{\rm \cite{CP}}
Let ${\bf C}$ be a $\dagger$C-theory. By a(n) \em (non-degenerate) observable \em for an object $X$ we mean any  commutative isometric dagger Frobenius comonoid $(X,\delta,\epsilon)$.
\end{definition}

Elsewhere we have referred to these non-degenerate observables as \em basis structures \em \cite{CD} or \em classical structures \em \cite{CP}.  What this mathematical concept stands for exactly will be explained below. Their name is justified by the following  theorem.

\begin{theorem}\label{thm:CPV}{\rm \cite{CPV}}
In ${\bf FHilb}$, observables in the sense of Definition \ref{def:observable} on a Hilbert space ${\cal H}$ are in bijective correspondence with the orthonormal bases of ${\cal H}$.  More precisely, each (unordered) orthonormal basis $\{|i\rangle\}_i$ yields an observable $({\cal H},\delta,\epsilon)$ with
\[\left\{\begin{array}{l}
\delta:{\cal H}\to{\cal H}\otimes {\cal H}::|\,i\,\rangle\mapsto|\,ii\,\rangle\\
\epsilon:{\cal H}\mapsto\mathbb{C}::|\,i\,\rangle\mapsto 1
\end{array}\right.\,.\]
Conversely, all observables in the sense of Definition \ref{def:observable} arise uniquely in this manner.
\end{theorem}

Hence in ${\bf FHilb}$ an orthonormal basis can be equivalently defined as a commutative isometric dagger Frobenius comonoid.  The orthonormal basis is `encoded' as the linear map which \em copies \em the vectors of that basis together with the linear map which \em uniformly erases \em them.  Of course, in quantum theory observables correspond to rays spanned by an element of  an orthonormal basis rather than to the basis itself.  For a discussion of observables in the sense of Definition \ref{def:observable} within ${\cal W}{\bf FHilb}$ we refer the reader to \cite{CD}.

The definition of an observable in a $\dagger$C-theory has an equivalent, purely diagrammatic, incarnation which suffices for our purposes in this paper. We set:
 \begin{center}
\qquad$\delta$\ $=$\ \raisebox{-0.30cm}{\epsfig{figure=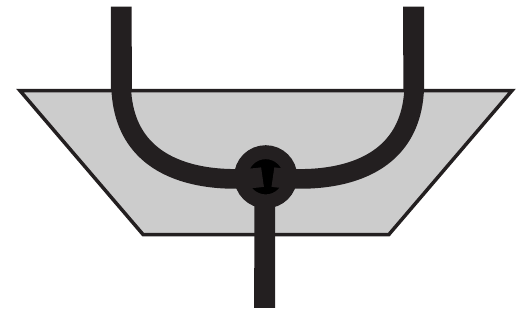,width=44pt}}\qquad\quad\qquad\qquad $\epsilon$\ $=$\ \raisebox{-0.30cm}{\epsfig{figure=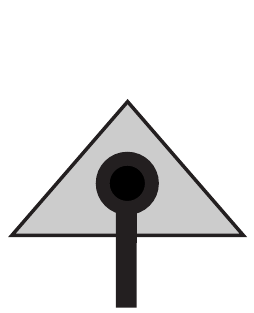,width=20pt}}
\end{center}

\begin{theorem}{\rm\cite{Kock,Lack,CPaq}} \label{spidertheorem}
All \underline{connected} diagrams built from `copying' ($\delta$), `erasing' ($\epsilon$), their daggers and straight wires, which have the same number of inputs and outputs, are equal. We represent these diagrams by a `spider':
\begin{center}
\epsfig{figure=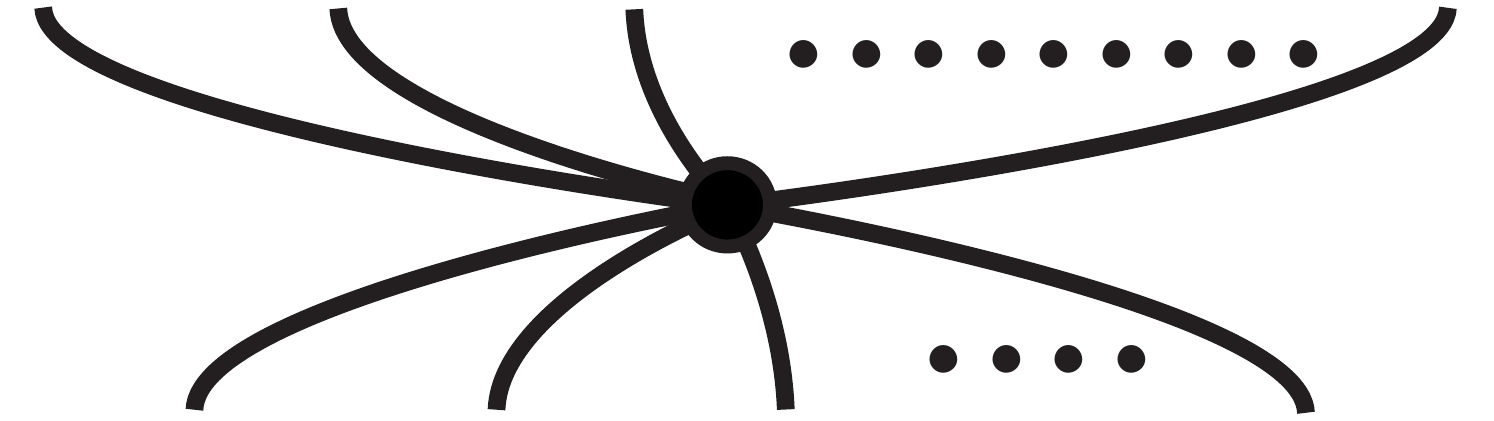,width=120pt}
\end{center}
Conversely, the defining equations of an observable in a $\dagger$C-theory (Definition \ref{def:observable}) are all implied  by the assumption that connected diagrams with the same number of inputs and outputs are equal. 
\end{theorem}

Observables on `subsystems' always lift to the whole system:

\begin{proposition}{\rm\cite{CD}}\label{prop:tensorobservable}
Two  observables $(A,\delta_X,\epsilon_X)$ and $(B,\delta_{X'},\epsilon_{X'})$ in a $\dagger$C-theory canonically induce an observable $(A\otimes B,\delta_{X\otimes X'} ,\epsilon_{X\otimes {X'}})$ with
\[
\delta_{X\otimes X'}=(1_A\otimes\sigma_{A,B}\otimes 1_B)\circ (\delta_X\otimes \delta_{X'}) \qquad\epsilon_{X\otimes {X'}}=\epsilon_X\otimes\epsilon_{X'}\,,
\]
where $\sigma_{A,B}$ is the morphism that swaps objects $A$ and $B$.  That is, diagrammatically,
 \begin{center}
\qquad$\delta_{X\otimes X'}$\ $=$\ \raisebox{-0.30cm}{\epsfig{figure=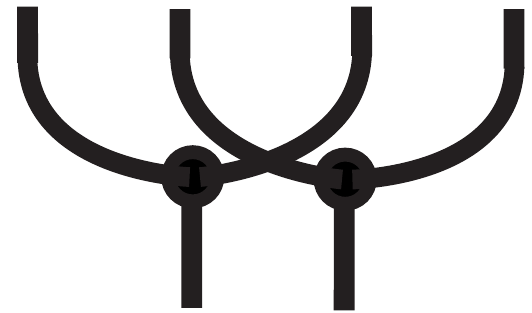,width=44pt}}\qquad\quad\qquad\qquad $\epsilon_{X\otimes {X'}}$\ $=$\ \raisebox{-0.30cm}{\epsfig{figure=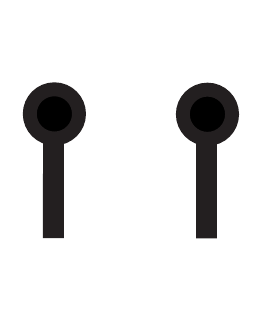,width=20pt}}
\end{center}
\end{proposition}

\begin{proposition} {\rm\cite{CP}}
Each observable  $(A,\delta,\epsilon)$ in $\dagger$C-theory induces a self-dual dagger compact structure $(A,\eta:=\delta\circ\epsilon^\dagger: \II\to A\otimes A)$, diagrammatically,
\[
\raisebox{-0.30cm}{\epsfig{figure=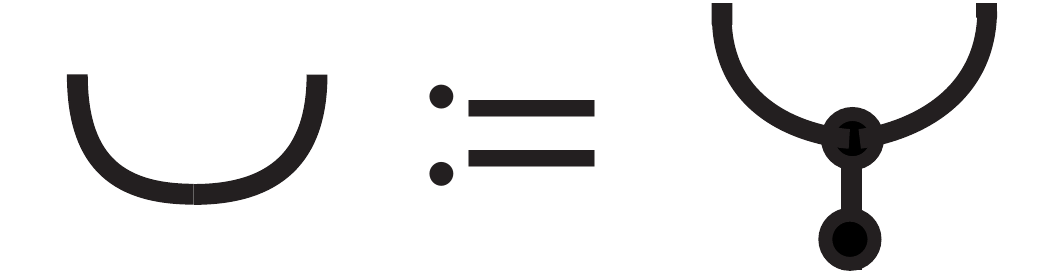,width=95pt}}\ ,
\]
where `compactness' means that:
\[
(\eta^\dagger\otimes 1_A)\circ(1_A\otimes\eta)=1_A\qquad\quad \sigma_{A,A}\circ\eta=\eta\,,
\]
that is, diagrammatically,
\[
\raisebox{-0.30cm}{\epsfig{figure=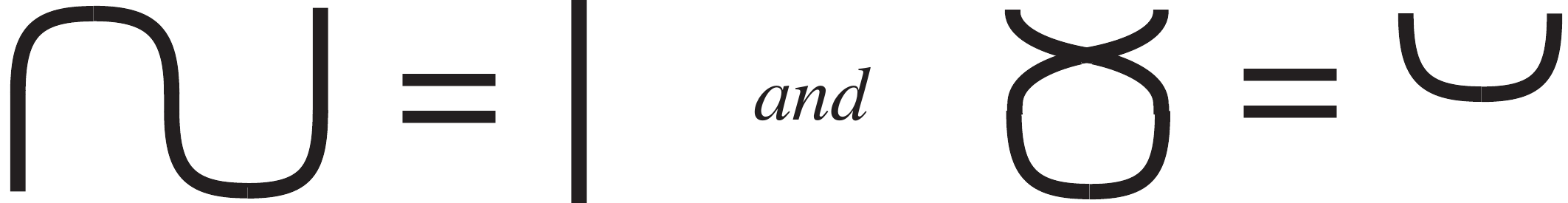,width=200pt}} \ .
\]
\end{proposition}

Given two such induced compact structures $(A,\eta_A)$ and $(B,\eta_B)$, and an arbitrary morphism $f:A\to B$, we can define abstract notions of the \em transpose \em morphism $f^{*}:B\to A$ and the \em conjugate \em morphism $f_{*}:B \to A$ \cite{AC}, respectively as follows
\[
f^*:=(\eta^\dagger_B\otimes 1_A)\circ(1_B\otimes f\otimes 1_A)\circ (1_B\otimes\eta_A)
\]
\[
\ \ f_*:=(\eta^\dagger_A\otimes 1_B)\circ(1_A\otimes f^\dagger\otimes 1_B)\circ (1_A\otimes\eta_B)\,.
\]
diagrammatically,
\[
f^* := \raisebox{-0.47cm}{\epsfig{figure=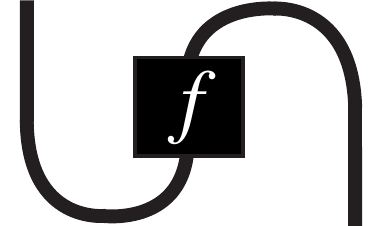,width=54pt}}
\qquad\qquad\qquad f_*:= \raisebox{-0.47cm}{\epsfig{figure=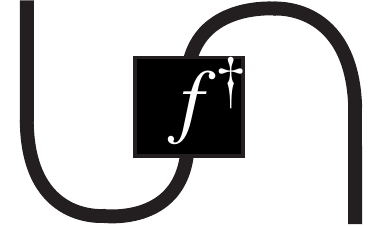,width=54pt}}
\]
 We also refer to a dagger compact structure $(A,\eta)$ as a \em Bell state\em. A graphical interpretation of Bell states can be found below in Definition \ref{def:GHXstate}.

Let $\lambda_\II: \II\simeq \II\otimes\II$. Now we define abstract counterparts of the basis vectors which are copied in \textbf{FHilb}:

\begin{definition} {\rm\cite{CD}} \label{eigenstate}
The \em eigenstates \em of an observable $(X, \delta, \epsilon)$ in a $\dagger$C-theory are all states $x:\II\to X$ which satisfy $\delta\circ x=(x\otimes x)\circ\lambda_\II$, $\epsilon\circ x=1_\II$ and $x_*=x$,  that is,
\[
\raisebox{-0.3cm}{\epsfig{figure=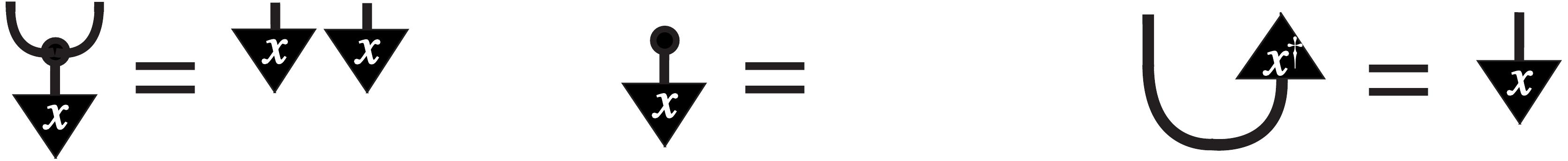,width=330pt}} \ .
\]
\end{definition}

The first of these conditions tells us that `eigenstates commute through dots'. \blk
Eigenstates moreover lift from subsystems to the whole system:

\begin{proposition}{\rm\cite{CD}}\label{prop:tensorobservableBIS}
If $x$ is an eigenstate for observable $(A,\delta_X,\epsilon_X)$ and $x'$ is an eigenstate for observable $(B,\delta_{X'},\epsilon_{X'})$ then $x\otimes x'$ is an eigenstate for  $(A\otimes B,\delta_{X\otimes {X'}} , \epsilon_{X\otimes {X'}})$ as defined in Proposition \ref{prop:tensorobservable}.
\end{proposition}

\subsection{GHZ states in $\dagger$C-theories}

\begin{definition}\label{def:GHXstate}
A \em GHZ structure \em for an object $X$ in a $\dagger$C-theory  is a triple
\[
\left(  X\ {\bf,}\  \Psi: \II\to   X\otimes  X\otimes  X\ {\bf ,}\  \epsilon:  X\to \II\right)
\]
where $\Psi$ is called a GHZ \em state\em, with
\bit
\item $\Psi$ symmetric i.e.
\[
\epsfig{figure=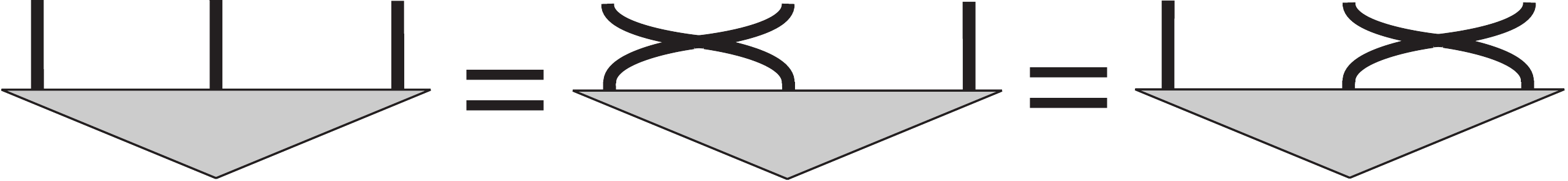,width=200pt}
\]
\item $(\epsilon\otimes 1_{X\otimes X})\circ\Psi$ is a Bell state i.e 
\[
\epsfig{figure=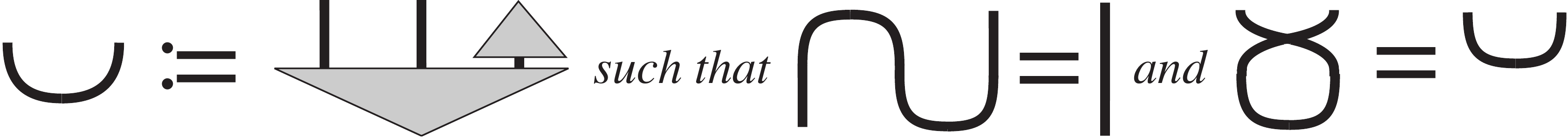,width=290pt}
\]
\item $\Psi$ and $\epsilon$ are both \em self-conjugate, \em i.e. 
\[
\epsfig{figure=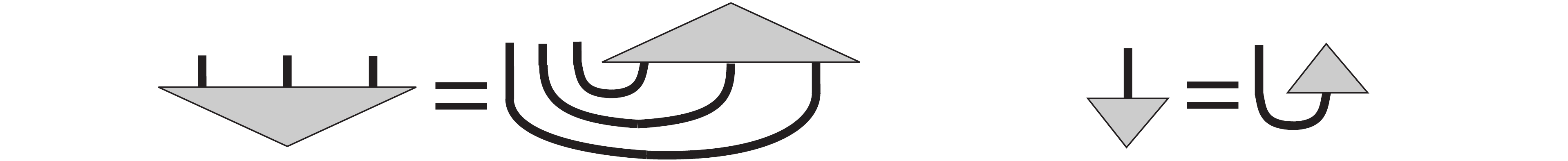,width=336pt}
\]
\item when `tracing out' two subsystems we obtain the maximally mixed state:
\[
\hspace{4cm}\epsfig{figure=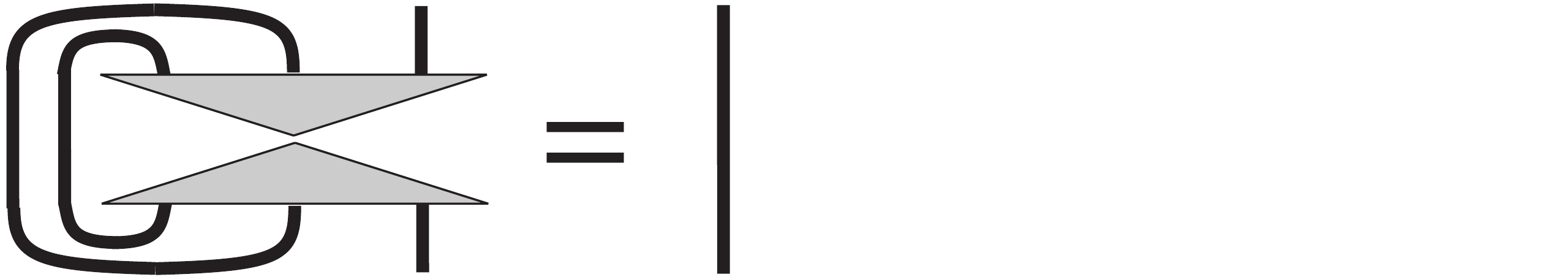,width=224pt}
\]
\eit
\end{definition}

\begin{theorem}\label{thm:GHZobservablecorrespondence}
GHZ structures in a $\dagger$C-theory  are in bijective correspondence with observables in that $\dagger$C-theory via the correspondence:
\[
\epsfig{figure=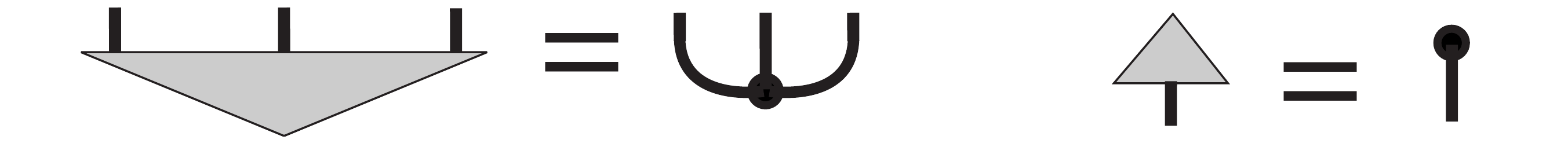,width=240pt}
\]
which assigns to each observable a GHZ structure, and its converse:
\[
\epsfig{figure=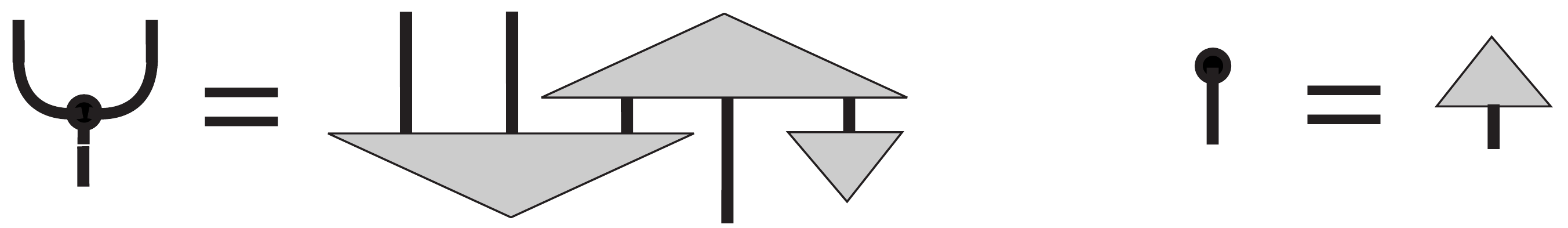,width=240pt}
\]
which assigns to each GHZ structure an observable.
\end{theorem}
\bpf
This can straightforwardly be verified using Theorem 16.2 in \cite{CP}.
\endproof\newline

\subsection{Phase groups in $\dagger$C-theories}

\begin{proposition} {\rm\cite{CD}}
For $(X,\delta,\epsilon)$ an observable in a $\dagger$C-theory let \vspace{2mm}
\bit
\item $states_X:=\{x:\II\to X\}$\vspace{2mm}
\item $x\odot y:=\delta^\dagger\circ(x\otimes y)=\raisebox{-0.40cm}{\epsfig{figure=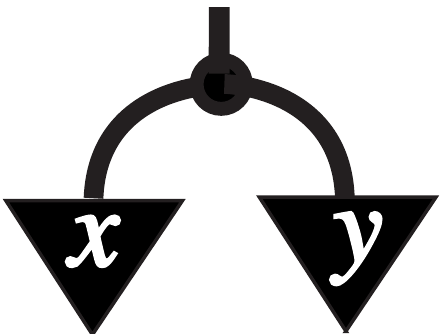,width=40pt}}$\vspace{-2mm}
\item $actions_X:=\{U_x:=\delta^\dagger\circ(x\otimes 1_X)=\raisebox{-0.40cm}{\epsfig{figure=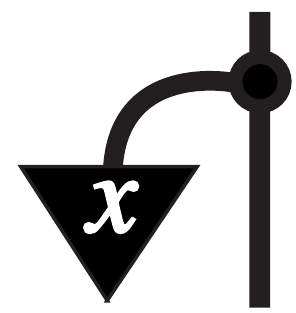,width=28pt}}\mid x:\II\to X\}$\vspace{-2mm}
\eit
\par\vspace{2mm}\noindent
  then $(states_X, \odot,\epsilon)$ and $(actions_X, \circ, 1_X)$ are isomorphic commutative monoids.
\end{proposition}

This follows straightforwardly from Theorem \ref{spidertheorem} (the spider theorem). 

\begin{example}
In ${\bf FHilb}$, for vectors $\psi=(\psi_1 , \ldots , \psi_n)$ and $\phi=(\phi_1 , \ldots , \phi_n)$  in the basis corresponding to $({\cal H}, \delta,\epsilon)$ via Theorem \ref{thm:CPV}, we have
$\psi\odot\phi=(\psi_1\cdot\phi_1, \ldots, \psi_n\cdot\phi_n)$, that is, $\psi \odot \phi$ is the component-wise product of $\psi$ and $\phi$.
\end{example}

\begin{proposition} {\rm\cite{CD}}
In ${\bf FHilb}$ a state $\psi$ (normalised so that $|\psi|^2=dim({\cal H})$) is unbiased with respect to the orthonormal basis corresponding to $({\cal H}, \delta,\epsilon)$ via Theorem \ref{thm:CPV} iff
$\psi_*\odot\psi=\epsilon^\dagger$.
\end{proposition}

Returning now to a general $\dag$C, let $dim(X)=\epsilon\circ\epsilon^\dagger$ for observable $(X, \delta,\epsilon)$.

\begin{definition} {\rm\cite{CD}}
In any $\dagger$C-theory a state $\psi:\II\to X$ with  $\psi^\dagger\circ\psi=dim(X)$ is \em unbiased \em for  an observable $(X, \delta,\epsilon)$  iff
\[
\psi_*\odot\psi=\epsilon^\dagger\qquad\mbox{that is}\qquad \raisebox{-0.40cm}{\epsfig{figure=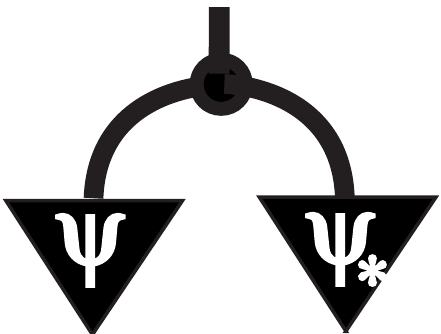,width=40pt}}=\raisebox{-0.30cm}{\epsfig{figure=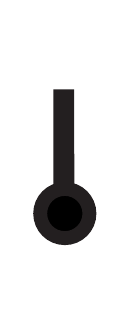,width=10pt}}\,.
\]
\end{definition}

By choosing $\psi^\dagger\circ\psi=dim(X)$ rather than $\psi^\dagger\circ\psi=1_\II$ as our normalisation convention we substantially simplify the expressions in this paper.  We refer to states $\psi:\II\to X$ which satisfy  $\psi^\dagger\circ\psi=dim(X)$ as \em states of length $\sqrt{dim(X)}$\em.

\begin{definition}  \cite{CD} \label{defmuo}
Two observables are \em mutually unbiased \em if the eigenstates of one are unbiased for the other.
\end{definition}

The diagrammatic significance of this definition is studied in detail in \cite{CD}.

\begin{theorem}\label{thm:phasegroup} {\rm\cite{CD}}
Let now
\bit
\item $U\mbox{\rm-}states_X$ be all states in $states_X$ unbiased with respect to the observable $(X,\delta,\epsilon)$
\item $U\mbox{\rm-}actions_X$ be all unitary  actions  in $actions_X$
\eit
then $(U\mbox{\rm-}states_X, \odot,\epsilon)$ and $(U\mbox{\rm-}actions_X, \circ, 1_X)$ are isomorphic  abelian  groups.  For $U\mbox{\rm-}states_X$ the inverses are provided by the adjoint, and for $U\mbox{\rm-}actions_X$ the inverses are provided by the conjugates for the induced compact structure.
\end{theorem}

\begin{definition} {\rm\cite{CD}}
We call the isomorphic groups of Theorem \ref{thm:phasegroup} the \em phase group\em.  \label{def:phasegroup}
\end{definition}

\begin{example}
In the case of a qubit the phase group is the circle of `relative phases'.  Concretely, when expressed in the standard basis, the unbiased states and the unitary actions have respective matrices:
\[
\psi_\alpha= \left(
 \begin{array}{c}
1\\ e^{i\alpha}
  \end{array}
\right)\qquad\qquad
U_\alpha=\delta\circ(\psi_\alpha\otimes 1_{\cal Q})= \left(
  \begin{array}{cc}
1&0\\0&e^{i\alpha}
  \end{array}
\right)\,.
\]
\end{example}

\subsection{GHZ correlations  in $\dagger$C-theories}\label{GHZcorrelationsection}

In Theorem \ref{thm:GHZobservablecorrespondence}, we showed the correspondence between observables and GHZ states. It comes as no great surprise then, that the measurement correlations of our GHZ states are closely related to the phase groups described in the previous section.

\begin{definition}
Let $(X, \delta, \epsilon)$ be an observable in a $\dagger$C-theory and let $(X, \Psi, \epsilon)$ be the corresponding GHZ state.  By a \em GHZ correlation triple \em we mean a triple $(x,x',x'')$ of states $x,x',x'':\II\to X$ of length $\sqrt{dim(X)}$ which is such that
\[
x''=(x\otimes x'\otimes 1_X)^\dagger\circ\Psi=\raisebox{-0.40cm}{\epsfig{figure=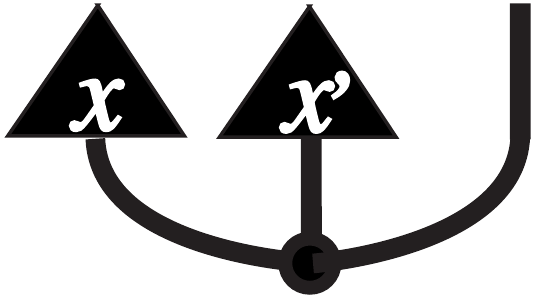,width=56pt}}\,.
\]
By \em GHZ correlations \em we mean the set
\[
\Gamma\subseteq{\bf C}(\II,X)\times {\bf C}(\II,X)\times{\bf C}(\II,X)
\]
consisting of all GHZ correlation triples.
\end{definition}

We can interpret these GHZ correlation triples in operational terms: when, in a measurement of the first and second qubit of the GHZ state $\Psi$, the effects $x^\dagger$ and $x'^\dagger$ occur then the third qubit is necessarily in state $x''$. If $x''=0$ this means that effects $x^\dagger$ and $x'^\dagger$ can never occur together.

\begin{proposition}\label{Prop:corr}
For GHZ correlations $\Gamma$ we have:
\bit
\item[{\bf i.}] For states $x,x',x'':I\rightarrow X$,  $(x,x';x'')\in \Gamma$ iff  $x''_*=x\odot x'$; in other words, correlation triples are exactly all triples of the form $(x,x',(x\odot x')_*)$ where $x,x',(x\odot x')_*\ne 0$.
\item[{\bf ii.}]  If $(\psi,\psi',\psi'')$ is a correlation triple and $\psi,\psi',\psi''$ are in the phase group then any triple obtained by permuting $\psi$, $\psi'$and $\psi''$ is also a correlation triple.
\eit
\end{proposition}
\bpf
Part {\bf i.} follows from (reversed triangles are the transposed):
\begin{center}
\centering{\epsfig{figure=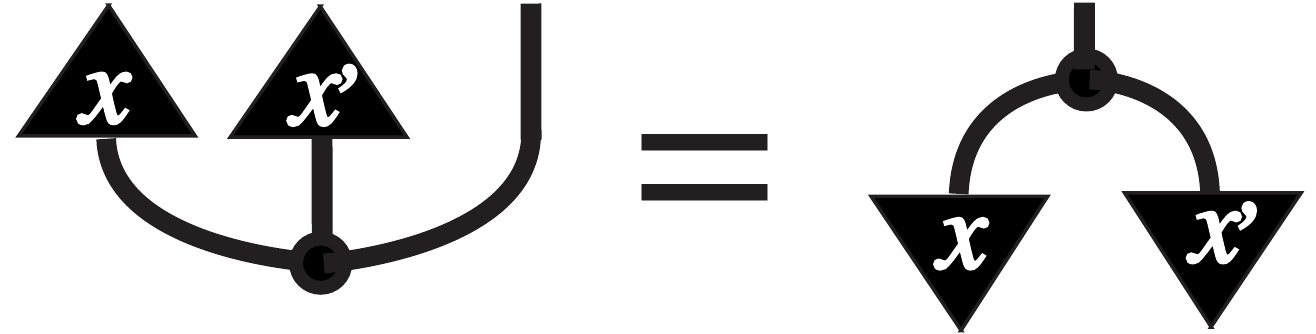,width=125pt}}
\end{center}
Part {\bf ii.} We have  $\psi'=(\psi_*\odot \psi)\odot \psi'=\psi_*\odot (\psi\odot \psi')=(\psi\odot(\psi\odot \psi')_*)_* $ so
$(\psi,(\psi\odot \psi')_*; \psi')= ( \psi , (\psi\odot \psi')_* ;  (\psi\odot(\psi\odot \psi')_*)_* )$ is indeed a correlation triple by part {\bf i} of this proposition.
\endproof\newline

\section{The key examples: \textbf{Stab} and \textbf{Spek}}

Having surveyed our unifying categorical framework we now proceed to consider two specific examples. The first is stabiliser qubit QM, a restricted version of standard qubit QM. The second is Spekkens's toy theory, which closely models many features of stabiliser QM, despite being essentially a local hidden variable theory. When considered within the categorical framework the similarity between the two is striking; and the precise difference between the two can be clearly pin-pointed. Furthermore the difference is to be found precisely in a certain categorical structure which is intimately involved in describing the physical phenomena where the two theories differ most significantly - locality v. non-locality.

\subsection{Stabiliser qubit quantum mechanics}

This is a subset of standard QM. The only systems in the theory are qubits, or collections of qubits. The states which these `qubits' can occupy are the stabiliser states of standard QM (these are the +1 eigenstates of tensor products of Pauli operators). For the single qubit there are six such states, the standard $\ket{0}, \ket{1}, \ket{+}, \ket{-}, \ket{i}$ and $\ket{-i}$. For two qubits we have all 36 possible tensor products of these single qubit states, plus 24 maximally entangled states, all related to the Bell state $\frac{1}{\sqrt{2}}\ket{00} + \ket{11}$ by local unitary operations. For three qubits we have many more states, including the GHZ state $\frac{1}{\sqrt{2}}\ket{000} + \ket{111}$.

The time evolution of states is given by those unitary operations which preserve stabiliser states. Such operations are called \emph{Clifford unitaries} and form a group. In fact, all $n$-qubit Clifford operations can be simulated using the CNOT gate (which is itself a Clifford unitary), and the single qubit Clifford unitaries. These single qubit operations themselves form a group, isomorphic to the permutation group $S_4$. The only measurements allowed in the theory are projective Pauli measurements.

Though a restricted version of QM, qubit stabiliser theory exhibits most of the key features of full QM. It has incompatible observables. There is a no-cloning theorem. Local hidden variable no-go proofs hold, as we shall soon see (although in the case of stabiliser QM we need to employ three qubit states, as in the GHZ no-go proof: although we have the Bell state, making Pauli measurements alone cannot violate Bell inequalities).

We have chosen to investigate stabiliser QM rather than the full theory, because it is much closer to the second theory which we will consider.

\subsection{Spekkens's toy theory}

We don't have space here to give full details of Spekkens's toy theory, these can be found in Ref.~\cite{Spekkens}. A brief description of the key points will suffice. The theory attempts to approximate stabiliser qubit QM: there is only one type of system, which is something like a qubit, and the states are discrete. The theory does not employ vector space. Instead a single system is described by a four state phase space. The actual state occupied in the phase space is called the \emph{ontic} state. However, the theory posits a fundamental restriction on our knowledge of the ontic state. This restriction is the fundamental principle of the theory, called the \emph{`knowledge balance principle'}. In full generality this principle is a bit awkward to state, but in the case of a single `qubit' it boils down to saying that we can at best know that the system is in one of two ontic states, \emph{with equal probability}. Our state of knowledge - the \emph{epistemic} state - is the toy theory's analogue of the quantum state. The theory is clearly, by construction, a local hidden variable theory.

Because of the equal probability caveat, mathematically the epistemic states of the `qubit' system are subsets of a four element set, hence there are six such states, just as in the case of stabiliser qubit QM. Invoking the knowledge balance principle, one can go on to derive the allowed states of composite systems, and all the operations on systems which are allowed in the theory. There turns out to be a one-to-one correspondence between the states and operations of the toy theory and those of the stabiliser theory, although how the operations combine together is not homomorphic. The operations of the toy theory transform between subsets of sets which represent the phase spaces of the various systems - thus they are most naturally described by \emph{relations} on these sets.

\subsection{The $\dagger$C-theories Stab  and Spek}

We now express both these theories within our $\dagger$C-theory framework. Interestingly, both theories can be defined in a constructive fashion:

\begin{definition} [${\bf Stab}$]\label{StabDef}
The $\dagger$C-theory ${\bf Stab}$ is the sub-$\dagger$C-theory of ${\bf FHilb}$ (recall example \ref{FHilbexample}) generated by:
\bit
\item   $n$th tensor powers of qubits ${\cal Q}:=\mathbb{C}^2$
\item   the single qubit Clifford unitaries
\item the linear map
\[
\delta_{stab}: {\cal Q}\to {\cal Q}\otimes {\cal Q}::\left\{\begin{array}{l}
|0\rangle\mapsto|00\rangle\\
|1\rangle\mapsto|11\rangle
\end{array}\right.
\]
together with the (necessarily unique) counit of this comultiplication, $\epsilon_{Stab}$.
\eit
\end{definition}

That this collection of operations is enough to generate all the states and operations of the stabiliser theory can be seen as follows:

\begin{itemize}
\item The Hadamard operation $H$ is a single qubit Clifford unitary.
\item ${\sf CNOT} := (1_{\cal Q}\otimes (H\circ\delta_{Stab}^\dagger\circ(H\otimes H)))\circ(\delta_{Stab}\otimes 1_{\cal Q})$
\item Arbitrary $n$-qubit Clifford unitaries $U_{\textrm{Clifford}}$ can be generated from the single qubit Clifford unitaries and ${\sf CNOT}$.
\item An arbitrary $n$-qubit stabiliser state $\Psi_{\textrm{stabilizer}} = U_{\textrm{Clifford}} (\epsilon^{\dagger}\otimes\epsilon^{\dag}\otimes\dots\otimes\epsilon^{\dag})$
\end{itemize}

Note that a similar construction actually applies to \textbf{FHilb} if we substitute the single qubit unitaries for the single qubit Clifford unitaries.

It is straightforward to verify that $(Q, \delta_{Stab}, \epsilon_{Stab})$ is an \emph{observable} as defined in Section \ref{observablesection}. The abstract GHZ state derived via Theorem \ref{thm:GHZobservablecorrespondence} is exactly the standard GHZ state $\frac{1}{\sqrt{2}}\ket{000} + \ket{111}$, which, as mentioned earlier, is a stabiliser state. All the results of Section \ref{keyfeaturessection}, on phase groups, correlation triples etc. apply.

\begin{proposition}\em \label{StabisMUQT}
The object $Q$ in \textbf{Stab} has three observables in total: the one mentioned in Definition \ref{StabDef}, and two others which copy the vectors $\ket{+}$ and $\ket{-}$, and $\ket{i}$ and $\ket{-i}$ respectively. All three observables are mutually unbiased.
\end{proposition}

\bpf
That these are the only other observables on $Q$ follows as a corollary of Theorem \ref{thm:CPV}, and the fact that \textbf{Stab} is a sub-category of \textbf{FHilb}.
That they are all mutually unbiased follows from straightforward computation.
\endproof

\begin{definition}\cite{Spek}[${\bf Spek}$]
The $\dagger$C-theory  ${\bf Spek}$ is the sub-$\dagger$C-theory of ${\bf FRel}$ (recall example \ref{FRelexample}) generated by:
\bit
\item   $n$th powers of qubits $\four:=\{1,2,3,4\}$
\item   all permutations on $\four$
\item the relation
  \[
\delta_{Spek}:  \four\to\four\times\four::\left\{\begin{array}{l}
1\mapsto\{(1,1),(2,2)\}\\
2\mapsto\{(1,2),(2,1)\}\\
3\mapsto\{(3,3),(4,4)\}\\
4\mapsto\{(3,4),(4,3)\}
\end{array}\right.
\]
together with the (necessarily unique) unit of this comultiplication, $\epsilon_{Spek}$.
\eit
\end{definition}

That these relations are sufficient to generate all the states and operations of Spekkens's toy theory (and no more) is not at all obvious, and is proved in \cite{Spek}. Perhaps unsurprisingly, given our choice of notation,  $(\four,\delta_{Spek},\epsilon_{Spek})$ turns out to be an observable. All the results of Section \ref{keyfeaturessection}, on phase groups, correlation triples etc. again apply.

\begin{proposition}\em \label{SpekisMUQT}
The object $\four$ in \textbf{Spek} has three observables in total. All three observables are mutually unbiased.
\end{proposition}

\bpf
The three observables are detailed in \cite{Spek}. That these are the only observables is shown in \cite{Dusko}. That they are mutually unbiased follows from straightforward computation.
\endproof

\begin{remark}
The use of relations in our construction actually leads to something we would term a \emph{possibilistic} theory. The scalars in \textbf{FRel} and thus in \textbf{Spek} are the Booleans. Such a theory can't really tell us the probability of any measurement outcomes, only whether such outcomes are possible or not. This is actually adequate for our later discussions of non-locality, since the kind of non-locality proofs we will invoke only involve measurement probabilities of 0 and 1. However, it should be noted that there is a well-defined procedure for modifying \textbf{Spek} so that its scalars are positive real numbers, and we can discuss probabilities.
\end{remark}


\subsection{Pinpointing the difference between Spek and Stab}

Our definitions of \textbf{Stab} and \textbf{Spek} are in terms of concrete vector spaces and linear maps, sets and relations. This allows us to make a clear connection with the way in which the theories were originally formulated. From our categorical perspective however the internal structure of the objects of a category is irrelevant, only the algebra of composition of morphisms is important. From this perspective, both \textbf{Stab} and \textbf{Spek} are generated by:

\bit
\item $n$th powers of qubit objects $Q$
\item the group $S_4$ acting on $Q$
\item an observable: $\delta: Q\rightarrow Q\otimes Q$ and its unit $\epsilon:Q\rightarrow I$
\eit

By definition, we know that the $\delta$ and $\epsilon$ morphisms always combine in the same way: according to Theorem \ref{spidertheorem}. And by specifying the group $S_4$ we have ensured that the group elements combine with one another in the same way in both cases. From this point of view it looks like \textbf{Stab} and \textbf{Spek} might be the same theory viewed in abstract categorical grounds. But this can't be the case: they describe quite different physical theories!

In fact the difference lies in the way that the group elements interact with the observable. One key example of such an interaction is the phase group. And indeed it is straightforward to verify that the phase groups of the qubit observables of \textbf{Stab} and \textbf{Spek} differ:

\begin{theorem}
The phase group for qubits in  ${\bf Stab}$ is the four element cyclic group $Z_4$ and the phase group  for qubits in  ${\bf Spek}$ is the Klein four group $Z_2\times  Z_2$.
\end{theorem}
\bpf
Straightforward computation.
\endproof\newline

In the next section we will show that this mathematical difference between the theories is intimately related to one of their key physical differences: the presence or absence of non-locality.

\section{Mutually unbiased qubit theories}

We have mentioned how Spekkens's toy theory and stabiliser qubit QM are similar kinds of theory: in both cases there is a discrete collection of states; in both cases the `qubit' system's observables (of which there are three) are all mutually unbiased. We next try to formally pin down the features which these theories share, within our categorical framework.

\begin{definition}
A \emph{mutually unbiased qubit theory}, or MUQT, is a dagger symmetric monoidal category with basis structures, which satisfies the following additional conditions:

\begin{enumerate}

\item The objects of the category are $I$, $Q$ (which will represent a qubit-like system), and $n$-fold tensor products of $Q$, i.e. $Q\otimes Q\otimes\dots\otimes Q$.

\item The observables on any given object are all \emph{alike}: that is to say, they have the same number of eigenstates, and the same phase groups.

\item The observables of $Q$ are all mutually unbiased (recall Definition \ref{defmuo}).

\item All states of $Q$ (i.e. morphisms of type $I\rightarrow Q$) are eigenstates of some observable.

\item $Q$ has \emph{three} observables, each with \emph{two} eigenstates.

\end{enumerate}

\end{definition}

Various results follow directly from this definition. (iv) and (v) together imply that $Q$ has six states. (iii) and (iv) together imply that, with respect to any observable on $Q$, all states are either eigenstates or unbiased. We can further conclude that each observable on $Q$ has two eigenstates and four unbiased states.

\begin{proposition}\em
\textbf{Stab} and \textbf{Spek} are both MUQTs.
\end{proposition}

\bpf
This follows from the definitions of the categories, and Propositions \ref{StabisMUQT} and \ref{SpekisMUQT}.
\endproof\newline


\subsection{Classification}

We will show that in a MUQT the possibilities for the basis structures on $Q$ are quite limited. More precisely the GHZ correlations can take one of two forms, and \textbf{Stab} and \textbf{Spek} cover these two possibilities.

The outline of this argument is fairly straightforward. Firstly, we recall the connection established in \ref{Prop:corr} between GHZ correlations and the monoid generated by the corresponding observable. We will shortly show that in a MUQT the monoid generated by the basis structures on $Q$ is completely determined by their phase group. Next we note that the phase group is an Abelian group, and has as many members as the basis structure has unbiased states, in this case four. Finally we recall that there are only two Abelian groups of four elements, the cyclic group $Z_4$ and the Klein four-group $Z_2 \times Z_2$.

So it simply remains to prove the first step, that in a MUQT the GHZ correlations on $Q$ are completely determined by the phase group. Recall Definition \ref{eigenstate}  of an eigenstate. From the axioms of an eigenstate it immediately follows that $x^\dagger\circ x=1_\II$.  More specifically, if $\delta\circ x=(x\otimes x)\circ\lambda_\II$ and $x_*=x$, then we have that $\epsilon\circ x=x^\dagger\circ x$.

\begin{lemma}\label{lm:zero1}
For $x,x':\II\to X$ eigenstates we have  $(x^\dagger\circ x')^2=x^\dagger\circ x'$.
\end{lemma}
\bpf
$(x^\dagger\circ x')^2=\lambda_\II^\dagger\circ(x\otimes x)^\dagger\circ(x'\otimes x') \circ\lambda_\II=x^\dagger\circ\delta^\dagger\circ\delta\circ x' =x^\dagger\circ x'$.
\endproof

\begin{lemma}\label{lm:zero2}
If for $x,x':\II\to X$ eigenstates $x^\dagger\circ x'=1_\II$ then $x=x'$.
\end{lemma}
\bpf
Ignoring natural isomorphisms,
$(1_X\otimes x^\dagger)\circ\delta\circ x'=(1_X\otimes x^\dagger)\circ(x'\otimes x')=
(x^\dagger\circ x')\cdot x'=x'$ (where we use $\cdot$ in place of $\otimes$ when the objects are numbers) from which it follows by $x_*=x$ that $x\odot x'=x'$.  By symmetry we also have
$x\odot x'=x$ and hence $x'=x\odot x'=x$.
\endproof\newline

Hence the inner product of two eigenstates is always an idempotent and for non-equal eigenstates this idempotent cannot be $1_\II$.

\begin{definition}
A $\dagger$C-theory  has a \em zero \em if it has exactly two idempotent numbers.
The idempotent number $0$ which is not the identity is referred to as \em zero\em.
\end{definition}

\begin{proposition}\label{prop:zero}
If two states $x\not=x'$ are eigenstates for an observable in a $\dagger$C-theory with zero then
we have $x^\dagger\circ x'=0$ and $x^\dagger\circ x=1_\II$.
\end{proposition}
\bpf
Follows from Lemma \ref{lm:zero1} and Lemma \ref{lm:zero2}.
\endproof\newline

In $\mathbb{R}$,  $\mathbb{R}_+$ and $\mathbb{C}$ the only idempotents are $0$ and $1$.  We will furthermore assume that any $0$-multiple of a state $\psi:\II\to A$ is a unique (trivial) state which we also denote by $0_A$.   We will indicate reliance on this assumption that there is a unique `absorbing idempotent number' by $(0)$.  This assumption is conceptually  justified by the interpretation of numbers as probabilistic weights -- see Example \ref{ex:wp}.

\begin{lemma}{\rm\cite{CD}}
For $x:\II\to X$ an eigenstate and $\psi:\II\to X$ unbiassed we have:
\[
dim(X)\cdot(x^\dagger\circ \psi)^\dagger\cdot(x^\dagger\circ \psi)=1_\II\,.
\]
\end{lemma}

Setting $|\langle x|\psi\rangle|^2:=(x^\dagger\circ \psi)^\dagger\cdot(x^\dagger\circ \psi)$ and assuming that $dim(X)$ admits an inverse $1/ D$, i.e.~$dim(X)\cdot 1/ D=1_\II$, results in the familiar form  $|\langle x|\psi\rangle|^2=1/ D$.  When we now subject a $\dagger$C-theory to the ${\cal W}$-construction of \cite{deLL} discussed in Example \ref{ex:wp}, then in the newly constructed category we have
\[
\langle {\cal W}x | {\cal W}\psi\rangle=({\cal W}x)^\dagger\circ {\cal W}\psi= (x^\dagger\circ \psi)^\dagger\cdot(x^\dagger\circ \psi)= 1/ D\,.
\]
We will assume below that we always are in a `$\dagger$C-theory without global phases' i.e.~a $\dagger$C-theory which is invariant under the ${\cal W}$- construction.   We will indicate reliance on this assumption by $({\cal W})$.  This assumption is again conceptually justified by the interpretation of numbers as probabilistic weights -- see Example \ref{ex:wp}.

\begin{remark}
Note that while ${\bf Spek}$, as a subcategory of ${\bf FRel}$, obviously has no global phases, it does have non-trivial relative phases, namely $Z_2\times  Z_2$.
\end{remark}

\begin{lemma}\label{lem:determine}
Let $(X, \delta, \epsilon)$ be an observable in a $\dagger$C-theory, let $\psi,\phi:\II\to X$ be unbiased for it and let $x\not=x':\II\to X$ be eigenstates for it.  Then we have:
\ben
\item[{\rm(1)}] $x\odot x=x$ and $x\odot x'= (x^\dagger\circ x') \cdot x\stackrel{(0)}{=}0$\,;
\item[{\rm(2)}] $x\odot \psi= (x^\dagger\circ \psi)\cdot x\stackrel{({\cal W})}{=}1/D \cdot x$\,;
\item[{\rm(3)}] $\psi\odot\phi$ is completely determined by the phase group.
\een
\end{lemma}
\bpf
For (1) we have:
\[
\epsfig{figure=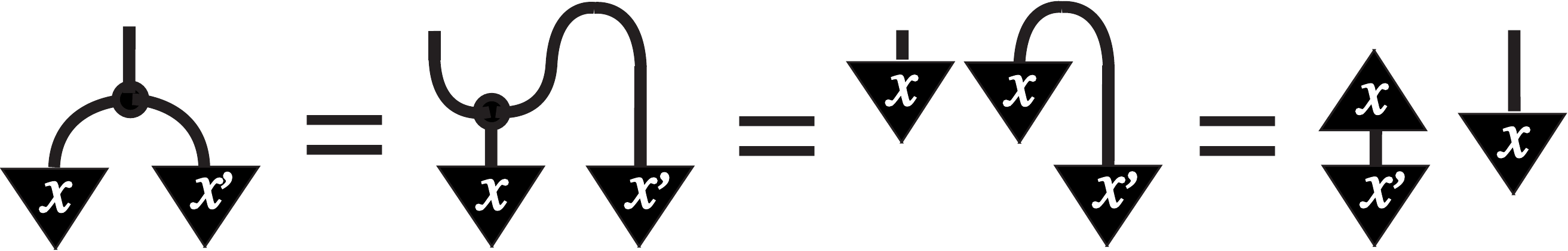,width=240pt}
\]
where the last step follows by $x=x_*$. Hence $x\odot x'=(x^\dagger\cdot x') \cdot x=0 \cdot x=0$.
If rather than $x'$ we would have considered $x$ itself then this graphical argument yields
$x\odot x=  (x^\dagger\cdot x) \cdot  x =x$.  For (2) the same graphical argument, now substituting $\psi$ for $x'$, results in $x\odot x' =(x^\dagger\circ \psi) \cdot x =1/D \cdot x$.  (3) is simply a consequence of the definition of the phase group (definition \ref{def:phasegroup}).
\endproof

\begin{corollary}\label{col:phasegroupclassefies}
Consider a $\dagger$C-theory which obeys  (0) and (${\cal W}$) and consider an observable in it for which all states on the underlying object are either eigenstates or unbiased.  Then, the choice of phase group constitutes the only degree of axiomatic freedom for how the multiplication $-\odot-$ of the observable acts on states.
\end{corollary}

Next we can use Proposition \ref{Prop:corr} to make the link to GHZ correlations:

\begin{lemma}\label{lem:determineBIS}
Let $(X, \Psi, \epsilon)$ be a GHZ state in a $\dagger$C-theory, let $\psi,\phi:\II\to X$ be unbiased for it and let $x\not=x':\II\to X$ be eigenstates for it.  Then we have:
\ben
\item[{\rm(1a)}] $(x,x;x)$ is a correlation triple\,;
\item[{\rm(1b)}] there are no correlation triples involving both $x$ and $x'$\,;
\item[{\rm(2)}] $(x,\psi;x)$  is a correlation triple\,;
\item[{\rm(3)}] all correlation triples involving at least two phase group elements are of the form $(\psi,\phi;(\psi\odot\phi)_*)$ -- which by Prop.~\ref{Prop:corr} {\bf ii.} includes permutations thereof.
\een
\end{lemma}

\bpf
Using Proposition \ref{Prop:corr}, each of these items follows from the similarly numbered item of lemma \ref{lem:determine}.
\endproof\newline

\begin{corollary}\label{phaseisall}
Consider a $\dagger$C-theory which obeys  (0) and (${\cal W}$) and consider a GHZ state in it for which all states on the underlying object are either eigenstates or unbiased.  Then, the choice of phase group constitutes the only degree of axiomatic freedom for the corresponding GHZ correlations.
\end{corollary}

Finally considering that in a MUQT the phase group must have four elements, and that there are only two four element groups $Z_4$ and $Z_2 \times Z_2$, we can state our main result:

\begin{theorem}
The GHZ correlations of the `qubit' object in a MUQT can take only two forms, corresponding to the two four-element groups, $Z_4$ (as in the case of \textbf{Stab}) and $Z_2 \times Z_2$ (as in the case of \textbf{Spek}).
\end{theorem}

We conclude that, whilst there is a vast number of possible MUQTs, their GHZ correlations can take only one of two forms, and \textbf{Stab} and \textbf{Spek} exemplify the two possibilities.

\subsection{Link to non-locality}

The GHZ correlations in the theories are of particular interest, because these correlations are invoked in one of the most elegant `no-go' proofs showing that quantum mechanics cannot be explained by a local realist theory. For the full details of this famous proof the reader is referred to \cite{MerminGHZ}.

Note the following key points:

\begin{itemize}

\item \emph{This no-go proof also applies to stabiliser theory}.
The proof begins with a GHZ state. The key ingredients are the probabilities of outcomes when we measure the variables $X\otimes X\otimes X$, $X\otimes Y\otimes Y$, $Y\otimes X\otimes Y$, and $Y\otimes Y\otimes X$. GHZ states and Pauli measurements both survive the restriction from full QM to the stabiliser theory, so the proof applies equally well in this case, i.e. it is impossible to model stabiliser theory with a local realist theory.

\item \emph{The key structural ingredients of the proof are all present in any MUQT}\footnote{In fact, in this section we restrict attention to \emph{probabilistic} and \emph{possibilistic} MUQTs, since these are the only $\dagger$-C theories where it makes sense to discuss locality/non-locality.}.
We can depict all the key ingredients of the proof in our categorical framework. Diagrammatically the relevant probabilities are given by:
\begin{center}
\epsfig{figure=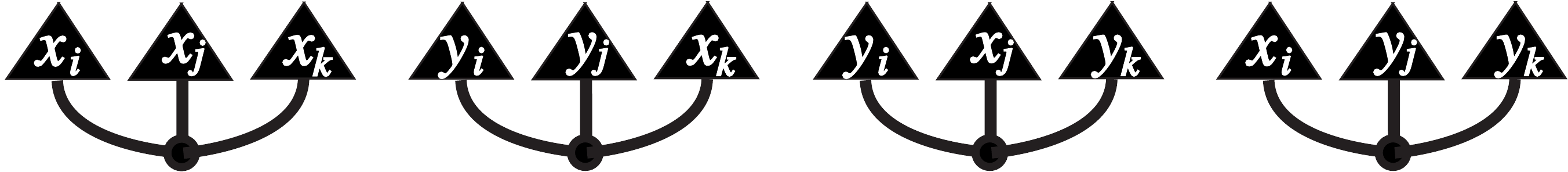,width=280pt}
\end{center}
In our abstract terminology we would say that the proof is employing a basis structure, and four of its unbiased states. An analogue of the argument could be reconstructed in any dagger symmetric monoidal category with these features, and with scalars which are numbers or Booleans. Certainly any MUQT will have an analogue of the proof, where the scalars pictured above are the GHZ correlations.

\item \emph{No $Z_4$ MUQT can have a local realist model}.
For MUQTs with a $Z_4$ qubit basis structure the proof will be identical to the quantum stabiliser case, ruling out a local realist model.

\item \emph{A local realist model can be constructed for the GHZ state in any $Z_2\times Z_2$ MUQT}. Hence, in the case of general MUQTs with $Z_2 \times Z_2$ correlations, we cannot rule out such a model, because we have a concrete example of a local realist theory, $\textbf{Spek}$, which exhibits exactly these correlations. Put another way, if we were presented with the data of a set of $Z_2 \times Z_2$ correlations, we could always explain them via the hidden variables of $\textbf{Spek}$.

\end{itemize}

Thus we can conclude that no MUQT of the $Z_4$ type can have a local realist interpretation, since at least one of its states (the GHZ) does not have such an interpretation. We cannot conclude that all MUQTs of the $Z_2 \times Z_2$ type will have a local realist interpretation, since they might have other states which had no such interpretation. We can at least conclude though, that GHZ-type no-go arguments will not work for them.

Turning this on its head, we can see that the $Z_4$ type basis structure, within our framework, is a structural fragment which embodies non-locality. If your theory has a basis structure of this type, then your theory has `got non-locality'. The $Z_2 \times Z_2$ structure has no non-locality. Whilst a $Z_2 \times Z_2$ type MUQT might have some other non-local piece of structure, the $Z_2 \times Z_2$ type basis structure cannot itself endow a theory with non-locality.

\section{Non-locality directly from abstract arguments} \label{groupnonlocalsection}

The arguments above are slightly round-about: we show that certain phase groups are exhibited by either \textbf{Stab} or \textbf{Spek}, which we know by other arguments to be non-local and local respectively. We then conclude that $Z_4$ GHZ states must have non-locality, whereas $Z_2\times Z_2$ GHZ states can not. In fact, in the case of $Z_4$ we can provide a more general argument directly from abstract reasoning.
\[
\xymatrix@=0.75in{
Z_4 \ar@{<->}[r] \ar@{<->}@/^1.0em/[rr]^{abstract\ \ \ \ } & {\bf Stab} \ar@{<->}[r] & \mbox{non-local}\hspace{-4mm}
}
\]
Let ${\cal O}_A$ be the set of observables on an object $A$ in a $\dagger$C-theory.  Let ${\cal E}_{o_A}$ be the set of eigenstates of an observable $o_A\in{\cal O}_A$.  We now define a notion of local realist representation which applies to arbitrary $\dagger$C-theories with $\mathbb{R}_+$ as numbers.  This can be extended to $\dagger$C-theories with more general numbers, as we show at the end of this section for the case of purely qualitative relational theories.

\begin{definition}\label{def:LHV}
Let ${\bf C}$ be a $\dagger$C-theory with $\mathbb{R}_+$ as numbers.  A state $\Psi:\II\to A_1\otimes\ldots\otimes A_n$ in ${\bf C}$ admits a \em local realist representation \em  if there exist:
\bit
\item a set of \em hidden states \em $\Xi\subseteq \prod_{o_{1}\in {\cal O}_{1}}{\cal E}_{o_{1}}\times\ldots\times \prod_{o_{n}\in {\cal O}_{n}}{\cal E}_{o_{n}}$ each of which assigns an eigenstate in ${\cal E}_{o_{i}}$ to each observable $o_{i}\in {\cal O}_{i}$ on each subsystem $A_i$, and we denote this eigenstate for $\xi\in\Xi$ by $\xi(o_{i})$
\item a $\sigma$-additive measure $\mu:{\cal B}(\Xi)\to\mathbb{R}_+$ with $\mu(\Xi)=1$
\eit
and these are such that for each choice of observables $o_{1}\in {\cal O}_{1}, \ldots, o_{n}\in {\cal O}_{n}$ and each choice of eigenstates $x_{1}\in {\cal E}_{o_{1}}, \ldots, x_{n}\in {\cal E}_{o_{n}}$  we have
\[
\mu\left( \{  \xi\in\Xi  \mid  x_{1}=\xi(o_{1}), \ldots,x_{n}= \xi(o_{n}) \}  \right)
=\left(x_{1}^\dagger\otimes\ldots\otimes x_{1}^\dagger\right)\circ\Psi\,.
\]
The $\dagger$C-theory ${\bf C}$ admits a \em local realist representation \em if each of its states admits a local realist representation
\end{definition}

We provide a no-go argument for GHZ states that applies to the GHZ states on qubits in  ${\bf Stab}$ and ${\bf FHilb}$.  This argument is not very different from the usual one \cite{MerminGHZ}, except for the fact that there is no reference to Hilbert space anymore and that a contradiction is directly drawn from the structure of the $Z_4$ phase group.

\begin{definition}
Let $(A,\Psi,\epsilon)$ be a GHZ state in a $\dagger$C-theory.  A \em forbidden triple \em is a triple of states
$(x,x';x'')$ such that  $x''$ and $x\odot x'$ are distinct eigenstates for the same observable.
\end{definition}

\begin{proposition}\label{prop:forbiddenzero}
If $(x,x';x'')$ is a forbidden triple for  GHZ state $(A,\Psi,\epsilon)$  in a $\dagger$C-theory with zero then we have $(x\otimes x'\otimes x'')^\dagger\circ\Psi=0$.
\end{proposition}
\bpf
Since $x''$ and $x\odot x'$ are distinct eigenstates for the same observable,
ignoring natural isomorphisms, we have $(x\otimes x'\otimes x'')^\dagger\circ\Psi= x''^\dagger\circ(x\otimes x'\otimes 1_A)^\dagger\circ\Psi=x''^\dagger\circ(x\odot x')=0$ by Proposition \ref{prop:zero}.
\endproof

\begin{theorem}
Let $(A,\Psi_Z,\epsilon_Z)$ be a GHZ state in a $\dagger$C-theory with $\mathbb{R}_+$ as numbers, which contains $Z_4$ as a subgroup of the phase group, and let the identity and the involutive element of this subgroup constitute the eigenstates of an observable $(A,\Psi_X,\epsilon_X)$, and its other two elements the eigenstates of an observable $(A,\Psi_Y,\epsilon_Y)$.  Then the state $\Psi_Z:\II\to A\otimes A\otimes A$ does not admit a local realist representation.
\end{theorem}
\bpf
We denote the identity of the phase group by $|+\rangle$ and the involutive element by $|-\rangle$, and the two other elements by $|\sharp\rangle$ and $|\!=\rangle$.  By the $Z_4$  structure we have:
 \[
|+\rangle\odot |+\rangle=|+\rangle\quad
|+\rangle\odot |-\rangle=|-\rangle\quad
|-\rangle\odot |-\rangle=|+\rangle\,.
\]
Hence, by Proposition \ref{Prop:corr} {\bf ii} we have that each correlation triple involving only states $\{|+\rangle,|-\rangle\}$ must have an even number of occurences of $|-\rangle$'s, and hence those with an odd number of $|-\rangle$'s are forbidden triples.  Also:
 \[
|\sharp\rangle\odot |\!=\rangle=|+\rangle\quad
|\!=\rangle\odot |\!=\rangle=|-\rangle\quad
|\sharp\rangle\odot |\sharp\rangle=|-\rangle\,.
\]
Hence, by Proposition \ref{Prop:corr} {\bf ii} we have that each correlation triple involving two states in $\{|\sharp\rangle,|\!=\rangle\}$ and one state in $\{|+\rangle,|-\rangle\}$ must have an odd number of occurrences of elements in $\{|-\rangle,|\!=\rangle\}$, and hence those with an even number of elements in $\{|-\rangle,|\!=\rangle\}$ are forbidden triples.  Assume that $\Psi$ admits a realist representation $(\Xi,\mu)$.  To distinguish between the three factors in $A\otimes A\otimes A$ we will denote them by $A_1, A_2, A_3$ respectively.  Using the notation of Definition  \ref{def:LHV},  we have for $o^{+/-}$ the observable with eigenstates $\{|+\rangle,|-\rangle\}$ that
\[
\mu\left( \left\{  \xi\in\Xi \Bigm|  x_{1}=\xi(o_{1}^{+/-}), x_{2}=\xi(o_{2}^{+/-}),x_{3}= \xi(o_{3}^{+/-}) \right\}  \right)=0
\]
whenever the number of $|-\rangle$'s in $(x_1,x_2,x_3)$ is odd by Proposition \ref{prop:forbiddenzero}. Hence
\[
\mu\left( \Delta_{odd}^{(1,2,3)}:=\left\{  \xi\in\Xi  \Bigm|  \mbox{odd $|-\rangle$'s in }
\left(\xi(o_{1}^{+/-}), \xi(o_{2}^{+/-}), \xi(o_{3}^{+/-})\right) \right\}  \right)=0\,.
\]
and so for $\Delta_{even}^{(1,2,3)}=\Xi\setminus\Delta_{odd}^{(1,2,3)}$ we have $\mu(\Delta_{even}^{(1,2,3)})=1$.
Similarly, for
\[
\Delta_{odd}^{(1)}:=\left\{  \xi\in\Xi  \Bigm|  \mbox{odd $|-\rangle$'s \& $|\!=\rangle$'s in }
\left(\xi(o_{1}^{+/-}), \xi(o_{2}^{\sharp/=}), \xi(o_{3}^{\sharp/=})\right) \right\}
\]
\[
\Delta_{odd}^{(2)}:=\left\{  \xi\in\Xi  \Bigm|  \mbox{odd $|-\rangle$'s \& $|\!=\rangle$'s in }
\left(\xi(o_{1}^{\sharp/=}), \xi(o_{2}^{+/-}), \xi(o_{3}^{\sharp/=})\right) \right\}
\]
\[
\Delta_{odd}^{(3)}:=\left\{  \xi\in\Xi  \Bigm|  \mbox{odd $|-\rangle$'s \& $|\!=\rangle$'s in }
\left(\xi(o_{1}^{\sharp/=}), \xi(o_{2}^{\sharp/=}), \xi(o_{3}^{+/-})\right) \right\}
\]
we have $\mu(\Delta_{odd}^{(1)})=\mu(\Delta_{odd}^{(2)})=\mu(\Delta_{odd}^{(3)})=1$
so $\mu(\Delta_{odd}^{(1)}\cap\Delta_{odd}^{(2)}\cap\Delta_{odd}^{(3)})=1$. It follows that there must be an odd number
of $|-\rangle$'s and  $|\!=\rangle$'s in
\[
\left(\xi(o_{1}^{+/-}), \xi(o_{2}^{\sharp/=}), \xi(o_{3}^{\sharp/=}),\xi(o_{1}^{\sharp/=}), \xi(o_{2}^{+/-}), \xi(o_{3}^{\sharp/=}), \xi(o_{1}^{\sharp/=}), \xi(o_{2}^{\sharp/=}), \xi(o_{3}^{+/-})\right).
\]
But due to the double occurrences of $\xi(o_{1}^{\sharp/=}),\xi(o_{2}^{\sharp/=}),\xi(o_{3}^{\sharp/=})$, this means an odd number of $|-\rangle$'s in $\left(\xi(o_{1}^{+/-}), \xi(o_{2}^{+/-}), \xi(o_{3}^{+/-})\right)$ so
$\Delta_{odd}^{(1)}\cap\Delta_{odd}^{(2)}\cap\Delta_{odd}^{(3)}\subseteq\Delta_{odd}^{(1,2,3)}$ and hence
$1=\mu(\Delta_{odd}^{(1)}\cap\Delta_{odd}^{(2)}\cap\Delta_{odd}^{(3)})\leq \mu(\Delta_{odd}^{(1,2,3)})=0$, hence a contradiction.
\endproof\newline

\section{Conclusions and further work}

We have described a categorical framework which is sufficiently flexible to accommodate both stabiliser QM, and Spekkens's toy theory,\footnote{While here we only considered the pure fragment of both theories, mixed states and operations can be straightforwardly adjoined by means of Selinger's CPM-construction \cite{Selinger}.} and which helps to cast light on the essential difference between the two. Structurally this difference is in the phase group: $Z_4$ in the case of \textbf{Stab} and $Z_2\times Z_2$ in the case of \textbf{Spek}. Physically the difference between the theories is that one is non-local whilst the other is local. We went on to show that it is the presence of the $Z_4$ phase group that makes stabiliser QM non-local. In fact, this structure suffices to show that full QM is non-local. 

We have furthermore defined a special class of toy theories, in which all `qubit' observables are mutually unbiased, which can all be modelled in the categorical framework. We have shown that the GHZ-correlations in these theories have phase groups $Z_4$ and $Z_2\times Z_2$.

We could extend the definition of a MUQT beyond `qubits', by allowing our basic system to have more observables, and its observables to have more eigenstates, while still insisting that the observables are all mutually unbiased. We would then have a more general \emph{mutually unbiased theory} or MUT. The result that the GHZ correlations in such a theory are completely determined by the phase group, established in Corollary \ref{phaseisall}, would still hold.

For example, in the case of qu\emph{trits}, there are four mutually unbiased observables, each with three eigenstates. Phase groups in this case would have nine elements. There are two nine-element groups, $Z_9$ and $Z_3 \times Z_3$. There is a well-defined way to extend stabiliser QM to higher dimensional systems and recently a `trit' version of Spekkens's toy theory was proposed \cite{SpekkensTris}. In this case the two theories coincide and their phase group is $Z_3 \times Z_3$. The toy theory is local by construction, and, as it turns out, so is the stabilizer formalism for qutrits. So is there a theory with phase group $Z_9$, and what kind of theory is it?  Solving questions of this kind is one avenue for future research.  Several other avenues suggest themselves, including:

\begin{itemize}

\item An obvious line of work beyond this is to consider `higher-dimensional' MUTs, beyond qubits and qutrits. The locality/non-locality properties of such theories will still be parametrised by Abelian groups. What sorts of locality/non-locality do we find? There is a well-defined way to extend stabiliser theory to any finite dimensional system. What is the phase group in each case? Can Spekkens's toy theory be extended beyond trits? What would its phase group be?

\item We have shown that the phase group is important in determining whether theories are local or exhibit quantum non-local correlations. In fact, theories have been proposed whose non-locality goes beyond that of quantum mechanics \cite{NLB}. Can these be accommodated within our framework? Some such theories are `qubit-like' in that they have two-valued observables. It would seem that mutually unbiased qubit theories are unable to exhibit super-quantum correlations, since with $Z_4$ and $Z_2\times Z_2$ we have exhausted the possibilities. Perhaps measurements on observables which are not mutually unbiased are required to display the super-quantum correlations.

\item We have shown in Section \ref{groupnonlocalsection} that there is an abstract argument that a $Z_4$ MUQT must be non-local. Could we construct a purely abstract argument that $Z_2 \times Z_2$ MUQTs must be local?
\[
\hspace{-18mm}\xymatrix@=0.75in{
\hspace{-2mm}Z_2\times Z_2 \ar@{<->}[r] \ar@{<->}@/^1.0em/[rr]^{\qquad ?\ \ \ } & {\bf Spek} \ar@{<->}[r] & \mbox{local}\hspace{-10mm}
}
\]
Could we develop a classification of groups, depending on whether they encode locality, quantum non-locality, or possibly super-quantum non-locality?

\end{itemize}


\end{document}